\documentstyle[12pt]{article}

\begin{document}                

\baselineskip=20pt
\newcommand{\beq}{\begin{equation}}
\newcommand{\eeq}{\end{equation}}
\newcommand{\beqa}{\begin{eqnarray}}
\newcommand{\eeqa}{\end{eqnarray}}
\newcommand{\bc}{\begin{center}}
\newcommand{\ec}{\end{center}}
\newcommand{\n}{\newline}
\newcommand{\ra}{\,\rangle\rangle}
\newcommand{\la}{\langle\langle\,}
\newcommand{\nn}{\nonumber\\}
\newcommand{\ov}{\overline}
\newcommand{\bi}{\bibitem}
\newcommand{\cl}{{\cal C}}
\newcommand{\gl}{{\cal G}}
\newcommand{\f}{{\bar f}}
\renewcommand{\theequation}{\thesection.\arabic{equation}}

\pagestyle{empty}
\bc
{\Large {
{\bf On the Out of Equilibrium Relaxation of the}
}}
\ec
\bc
{\Large {
{\bf Sherrington-Kirkpatrick model}
}}
\ec
\vspace{.3cm}
\bc
{\large {
L. F. Cugliandolo
\footnote{{\tt e-mail cugliandolo@roma1.infn.it}}
and
J. Kurchan
\footnote{{\tt e-mail kurchan@roma1.infn.it}}
}}
\ec
\vspace{.1cm}
\bc
{\large {
Dipartimento di Fisica,
}}
\linebreak
{\large {
Universit\`a di Roma I,
{\it La Sapienza},
}}
\linebreak
{\large {
I-00185 Roma, Italy
}}
\linebreak
{\large {
INFN Sezione di Roma I, Roma, Italy
}}
\linebreak
\ec
\vspace{.2cm}
\bc
{\large {
November, 1993
}}
\ec
\vspace{0.05cm}

\newpage

\bc
{\large {\bf Abstract}}
\ec

\vspace{.4cm}

We derive analytical results for the large-time relaxation of the
Sherrington - Kirkpatrick model in the thermodynamic limit, starting
from a random configuration.

The system never achieves local equilibrium in any fixed sector of
phase-space, but remains in an asymptotic out of equilibrium regime.

We propose as a tool, both numerical and analytical, for the study
of the out of equilibrium dynamics of spin-glass models the use of
`triangle relations' which describe the geometry of the configurations
at three (long) different times.

\vspace{3cm}

\newpage

\pagestyle{plain}
\setcounter{page}{1}
\setcounter{footnote}{1}

\section{Introduction}
\setcounter{equation}{0}

\vspace{.5cm}

In the past years most of the study of
spin-glass physics has been concentrated
on the Gibbs-Boltzmann equilibrium measure.
As a result of these efforts the mean-field theory is quite well understood
\cite{Pa,Mepavi}.
The picture that has emerged is one of a phase space with an
extremely complex landscape with many minima separated by
barriers, some of which are infinitely high.
Such divergent barriers lead
to ergodicity-breaking, a large  system is not able to explore
the phase-space in finite times.
For low dimensionalities mean-field is not exact, and the situation
is still controversial. In particular, the question of ergodicity breaking and
the existence of many pure states is still not settled \cite{Fihu,Mapari}.

One of the most striking phenomena  observed in the low temperature phase
of real spin glasses is the aging effect \cite{Lusvnobe,Vihaoc}:
the relaxation of the system depends on its history even after very long times.
Though aging effects seem unusual from the thermodynamical point of view, they
have been observed in numerous disordered systems, {\it e.g.}
in the mechanical properties of amorphous polymers \cite{St}, in
the magnetic properties of high temperature superconductors \cite{Ro}, etc.
The aging regime is an  essentially  out of equilibrium regime and
therefore
one has to face the dynamical problem
in order to understand most experiments;
the study of the Gibbs-Boltzmann weight yields only partial information.

Several phenomenological models have been proposed
to account for aging effects in spin glasses
\cite{others}.
In particular, a scenario for
the basic mechanism of aging to which we shall refer below
has been proposed by Bouchaud in Ref. \cite{Bo} (see also the
early work of Ref. \cite{Gi}).
The main idea is that of `weak ergodicity breaking', {\it i.e.}
the system is not allowed to access different ergodic components
but the relaxation takes place in a rough landscape
with `traps' and the distribution of the `trapping times'
does not have an upper bound.

Still one would like to have a satisfying microscopic description
of the dynamics of spin glasses and in particular of the effects
mentioned above. In this respect
there are, on the one hand,
some numerical simulations of realistic systems
\cite{Anmasv,Ri} that yield results in good agreement with experiments.
On the other hand, the analytical understanding of the
out of equilibrium relaxation {\it in the thermodynamic limit}, is
much less developed than that of the Gibbs-Boltzmann measure; it is only
recently that attention has been paid to this problem.

In Ref. \cite{Cuku} it was pointed out that mean-field models exhibit a rich
phenomenology in the out of equilibrium dynamics, qualitatively similar to
those of realistic models and experiments, and
some analytical results were obtained for a simple model, namely the
$p$-spin spherical model. One suggestion of that work is that
the out of equilibrium dynamics of mean-field models can be (at least
partially)
solvable analytically. The reason for this is the weakness of the long-term
memory: the system remembers all its past but in an averaged way,
short-time details tend to be washed away by the evolution.
Later, numerical analysis has shown that also the more standard
Sherrington-Kirkpatrick (SK) model captures the essential features
of aging phenomena \cite{Cukuri}.
Besides, Franz and M\'ezard have studied the out of equilibrium
relaxational dynamics of a particle in a random potential in infinite
dimensions.
They have numerically solved the closed set of mean-field causal equations
and have obtained with great precision results that are consistent
with the picture we shall assume here \cite{Frme}.

Some years ago, Sompolinsky and Zippelius introduced a
dynamical formalism for mean-field spin glasses
and used it to study the relaxation within an equilibrium state of the
SK model \cite{Sozi}.
Later, Sompolinsky
proposed to study the equilibrium (Gibbs-Boltzmann measure)
of spin-glasses by considering a relaxational dynamics after a very long
equilibration time and for large but finite $N$ \cite{So}.
The finiteness of $N$ guarantees ergodicity by allowing for the
penetration of barriers that diverge in the large $N$ limit.
The existence of divergent barriers led Sompolinsky to postulate a
hierarchical set of time scales which were taken as large
and eventually went to infinity with $N$.

However, in a true experimental situation the system is macroscopic and
it does not reach equilibrium even for very long times. Thus, in order to
make contact with the observations, we shall study
the relaxational dynamics starting from a random
configuration in the thermodynamic limit, {\em i.e.} making $N$ infinite from
the outset.
We shall concentrate on the asymptotics for large times, but
throughout this paper
we shall understand `large' as
$t \rightarrow \infty$ {\em after} $N \rightarrow \infty$ (the opposite
order to that considered by Sompolinsky).
Under these circumstances, the mean-field dynamical equations hold rigorously
and the solution is {\em unique} (if one considers the opposite order of
the limits, then one has to consider multiple solutions, just as in a
system with instantons \cite{Frku}).

Having already set $N \rightarrow \infty$,
the  mean-field equations of motion have no parameters that become infinite,
and {\em a priori} there are no
time scales that go to infinity with an external parameter.
However, the {\em solutions} to the mean-field equations of motion
may exhibit infinite time
scales under the conditions of weak ergodicity breaking.
In this scenario the landcape has, within an ergodic
component, finite barriers of {\em all heights}. The divergent barriers are
within this framework unsurmountable by hypothesis.
As the system ages,
it becomes more and more trapped, and it faces larger and larger barriers,
simply because it  has already crossed  the smaller ones and it has had more
time to find more `trapping' traps.
In other words an older system `sees' a more rugged landscape,
of course not because the actual landscape has changed with time, but because
of simple probabilistic reasons. The fact that there is no upper limit to
the size of the {\em finite} barriers makes it possible for this process
never to stop, and the system never to reach equilibrium.

Consider a two-time function, for example the autocorrelation function
$C(t_w+t,t_w)$. The preceding discussion suggests that the
behaviour of the relaxation of the correlation function in terms of
$t$ is affected by the overall `age' $t_w$.
Indeed, it turns out that the `age' $t_w$  automatically plays a
very similar role to the one played by $N$ in the Sompolinsky dynamics:
it controls the height of the barriers that are relevant
at such times. After a very large time $t_w$, the system has gotten itself
very trapped, and any subsequent motion (apart from a fast relaxation inside
a trap) takes times that blow-up with $t_w$.
It is at this point clear that if the age of the system
is what drives the time scales, then it is essential
that the correlation function be {\em non-homogeneous
in time} ({\it i.e.} not a function of $t$ exclusively).

Having not made the assumption of homogeneity in time
we shall obtain that
there is no time $t_{eq}$ such that for all
$t_1>t_2 > t_{eq}$ the two-time ($t_1,t_2$) functions
({\it e.g.} correlation and response functions)
obey the equilibrium relations,
Fluctuation-Dissipation Theorem (FDT) and homogeneity. This means that
the system does not reach equilibrium, not only in the (expected) sense
of not reaching the Gibbs-Boltzmann distribution, but in the wider sense
of not reaching any time-independent distribution in a fixed restricted sector
of phase space. In other words the dynamics is for all times something
different from local equilibrium.

One consequence of these assumptions is that for an infinite system
that is rapidly quenched below the critical temperature there is
no way to further change the external parameters slower than the
internal dynamics of the system, there being no upper time scale.
The question of adiabaticity becomes subtle at precisely the critical
temperature, where the upper time scales change from finite to infinite.
One expects that what happens in the adiabatic cooling of an
infinite system across the transition temperature is dependent of the
nature of the dynamical phase transition.

In short, we shall here analyze the out of equilibrium dynamics of the
SK model basically inspired by the previous results obtained for the simpler
$p$-spin spherical model \cite{Cuku}, the numerical simulations of Ref.
\cite{Cukuri}, and the phenomenological picture of Bouchaud \cite{Bo}.

\vspace{1cm}

The outline of the paper is as follows. In Section \ref{I} we review
the SK model and its relaxational dynamics. In Section \ref{II} we
present the assumptions of weak ergodicity breaking and weak long-term
memory on which the subsequent treatment is based. In Section \ref{III}
we discuss the asymptotic equations and their invariances and make two
further assumptions suggested by these invariances. Section \ref{IV} is
devoted to rather general properties of the geometry of the triangles
determined by the configuration at three different times and to the
discussion of the `correlation scales'. This last discussion is not
particular to the SK model. In Sections \ref{V} and \ref{VI} we construct the
solution to the asymptotic equations.

The results of this paper are by no means exhaustive, though they have
a few consequences that can be verified numerically and that are expressed
in terms of experimentally measurable quantites. Some of them are
presented in Section \ref{VII}  where we also mention some
results of Ref. \cite{Frme} relevant to this discussion.
In Section \ref{VIII} we
give a qualitative description of our results and contrast them
with those previously found in Ref. \cite{Cuku} for the $p$-spin
spherical model. Finally in the conclusions
we summarise our results, we discuss the relationship
with Sompolinsky dynamics and we point out some of the many open problems.

\vspace{.5cm}

\section{The relaxational dynamics of the SK model}
\label{I}
\setcounter{equation}{0}
\renewcommand{\theequation}{\thesection.\arabic{equation}}

\vspace{.5cm}

The SK Hamiltonian is $H =-\sum_{i < j}^N J_{ij} s_i s_j$ where
the interaction strenghts $J_{ij}$ are
independent random variables with a Gaussian distribution with zero mean
and variance
$\overline{ J_{ij}^2 }
=
1/(2 N)
$.
The overline stands for the average over the couplings.
The spin variables take values $\pm 1$. For convenience we consider a
soft-spin version
\beq
H =
-
\sum_{i < j}^N
J_{ij}
s_i s_j
+ a \sum_i (s_i^2-1)^2 +
\frac{1}{N^{r-1}}
\sum_{i_1 < \dots < i_r}^N
h_{i_1  \dots i_r}
s_{i_1} \dots s_{i_{r}}
\; ,
\nonumber
\eeq
$-\infty \leq s_i \leq \infty$, $\forall i$.
Letting $a \rightarrow \infty$ one recovers the Ising case, although this is
not essential.
 Additional
source terms ($h_{i_1 \dots i_{r}}$ time-independent)
have been included; if $r=1$ the usual coupling to a magnetic
field $h_i$ is recovered.

The relaxational dynamics is given by the Langevin equation
\beq
\Gamma_0^{-1} \, \partial_t s_i(t)
=
- \beta \frac{\delta H}{\delta s_i(t)}
+ \xi_i(t)
\; .
\label{lang}
\eeq
$\xi_i(t)$ is a Gaussian white noise
with zero mean and variance $2 \Gamma_0$.
The mean over the thermal noise is hereafter represented by
$\langle \;\cdot\;\rangle$.

The mean-field sample-averaged dynamics for $N \rightarrow \infty$
is entirely described by the evolution
of the two-time  correlation and the linear response functions \cite{Sozi}
\[
C(t,t')
\equiv
\frac{1}{N} \sum_{i=1}^N \overline { \langle s_i(t) s_i(t') \rangle }
\;\;\;\;
G(t,t')
\equiv
 \frac{1}{N}
\sum_{i=1}^N
\frac{\partial \overline {\langle s_i(t) \rangle}}
     {\partial h_i(t')}
\; .
\]

Following Ref. \cite{Cuku},
let us introduce the generalized susceptibilities
\beqa
I^r(t)
&\equiv&
\left.
\lim_{N\rightarrow\infty}
\frac{r!}{N^r}
\sum_{i_1 < \dots < i_r}
{
\frac{\partial
\overline{
<
s_{i_1}(t) \dots s_{i_r}(t)
>
}
}
{
\partial
h_{i_1 \dots i_r}
}
}\right|_{h=0}
\nn
&=&
r
\int_0^t dt' \, C^{r-1}(t,t') \, G(t,t')
\; ,
\label{int}
\eeqa
and their generating function
$P_d(q)$
\beq
\lim_{t \rightarrow \infty} \lim_{N \rightarrow \infty}
\left.
\left[ 1-
\frac{r!}{N^r}
\sum_{i_1 < \dots < i_r}
{
\frac{\partial
\overline{
<
s_{i_1}(t) \dots s_{i_r}(t)
>
}
}
{
\partial
h_{i_1 \dots i_r}
}
}\right|_{h=0}  \right]
=
\int_0^1 dq' \, P_d (q') \, {q'}^r
\; .
\label{aa}
\eeq
The interest of the function $P_d(q')$ is that if a system reaches
equilibrium within a fixed restricted sector of phase space then it
is easy to show that $P_d(q')$ should be just a delta function. We shall
see that this happens only for $T > T_c$.

\vspace{.5cm}

\section{Weak Ergodicity Breaking}
\label{II}
\setcounter{equation}{0}

\vspace{.5cm}

As in Ref. \cite{Cuku}, we shall make the following two assumptions,
supported by the numerical simulations of this model \cite{Cukuri}:

\vspace{1cm}

{\it i.} `Weak' ergodicity-breaking:
\beq
\lim_{t \rightarrow \infty} C(t,t')=0 \;\;\;\; \forall \;\mbox{fixed} \;\; t'
\; .
\label{web}
\eeq
This means that the system, after a given time $t'$, starts drifting away
(albeit slowly) until it reaches for sufficiently large times $t$ the maximal
distance $C=0$ (see section \ref{VII}).

This statement has to be slightly modified in the presence of a magnetic field;
in that case `$0$' has to be substituted by the  maximum distance compatible
with the remanent magnetization.
In particular this implies that the remanent magnetization $C(t,0)$
tends to  zero in the absence of a magnetic field.

\vspace{1cm}

{\it ii.}  `Weak' long-term memory:
\beq
\lim_{t \rightarrow \infty} \int_{0}^{t'} \; dt'' \;
G(t,t'')=\lim_{t \rightarrow \infty} \chi (t,t')=0 \;\;\;\; \forall \;
\mbox{fixed} \; t'
\; .
\label{wltm}
\eeq
$\chi (t,t')$ is the normalized (linear) response at time $t$
to a constant small magnetic field applied from $t'=0$ up to $t'=t'$,
often called the `thermoremanent magnetization'
(see section \ref{VII}).

This hypothesis is quite  crucial, since the response function
represents  the memory the system has of what happend at previous times: the
weakness of the long-term memory
implies that the system responds to its past in an averaged way, the details
of what takes place during a finite time tend to be washed away.

\vspace{1cm}

{\it iii.} Finally, we shall make the usual hypothesis that after a (long)
time $t'$ there is a quick relaxation in a `short'  time $t-t'$
to some value
$q$, followed by a slower drift away. The parameter $q$ is interpreted in
the Sompolinsky-Zippelius dynamics as the Edwards-Anderson parameter for
a state \cite{Sozi}. Here the word `state' certainly does not apply (since a
true state
is a separate ergodic component) but we may picture $q$ as the size of
a `trap' or the `width of a channel'. Within these traps the system is
fully ergodic while it becomes more and more difficult to escape a trap
as time passes.
The correlation and response functions are thus written in a way that
explicitly separates the terms corresponding to the relaxation within a trap:
\beqa
\begin{array}{rcl}
C(t,t')&=&C_{FDT}(t,t')+{\cal C}(t,t')
\; ,
\nn
G(t,t')&=&G_{FDT}(t,t')+{\cal G}(t,t')
\; .
\label{fdt}
\end{array}
\eeqa
Consistently, $C_{FDT}(t,t')$ and $G_{FDT}(t,t')$ are assumed
to satisfy the equilibrium relations, {\it i.e.} time homogeneity and the
fluctuation-dissipation theorem (FDT)
\beqa
\begin{array}{rcl}
C_{FDT}(t,t') &=& C_{FDT}(t-t')\nn
G_{FDT}(t,t') &=& G_{FDT}(t-t')
\end{array}
\;\;\;\;\;\;\;\;\;
 G_{FDT}(t-t')&=& \frac{\partial C_{FDT}(t-t')}{\partial t'} \; ,
\eeqa
and
\beqa
\begin{array}{rclrcl}
C_{FDT}(0)&=&1-q
\; ,
&
\;\;\;\;\;\;
\lim_{t-t' \rightarrow \infty} C_{FDT}(t-t') &=& 0
\; ,
\nn
{\cal C}(t,t) &=& q
\; ,
&
\lim_{t \rightarrow \infty} {\cal C}(t,t')&=&0
\; .
\end{array}
\eeqa

The equilibrium
dynamics within a state has been solved with these assumptions
\cite{Sozi}. Since this calculation remains the same for the out of
equilibrium dynamics, although the `state' must be reinterpreted
as a `trap', we shall not discuss it in this work.
We shall concentrate on the evolution of the long-time functions
${\cal C}$ and ${\cal G}$ and furthermore
we shall restrict ourselves to the dynamics
of the model at a temperature near and below the critical, $T=T_c-\tau$
with $\tau$ small.

\vspace{.5cm}

\section{Asymptotic Equations}
\label{III}
\setcounter{equation}{0}

\vspace{.5cm}

The full dynamical equations have been written down by
Sompolinsky and Zippelius \cite{Sozi}. They are rather
cumbersome because, just as in the static case, the
spin variables cannot be explicitly integrated away.

Under the assumptions made in the preceding section, {\it viz}
weak ergodicity breaking and weak long term memory,
one can find  equations for the
evolution of ${\cal C}$ and ${\cal G}$ {\em valid asymptotically
for large times} $t>t'$ near the transition.
They have been presented in Ref. \cite{Dofeio} and they
correspond to the dynamical counterpart of the `truncated model', the statics
of which was solved by Parisi \cite{Pa}. (The derivation can be
done in a way that makes clear the contact with the static free-energy
functional near $T_c$ by writing the dynamics in the supersymmetric
notation \cite{Ku,Cufrkume}). In the absence of a magnetic field
the equations read
\beqa
2 \, (\tau - q) \; {\cal C}(t,t')
+
y \, {\cal C}^3(t,t')
+
\int_{0}^{t'} \; dt'' \; {\cal C}(t,t'') {\cal G}(t',t'')
\;\;\;\;\;
& &
\nn
+
\int_{0}^{t'} \; dt'' \; {\cal G}(t,t'') {\cal C}(t',t'')
+
\int_{t'}^{t} \; dt'' \; {\cal G}(t,t'') {\cal C}(t'',t')
&=&
0
\; ,
\nn
\label{ceq}
\\
2 \, (\tau - q) \; {\cal G}(t,t')
+ \;
\int_{t'}^{t} \; dt'' \; {\cal G}(t,t'') {\cal G}(t'',t')
+
3y \, {\cal C}^2(t,t') \, {\cal G}(t,t')
&=&
0
\; ,
\nn
\label{geq}
\eeqa
$y=2/3$. In these equations causality is assumed
\beq
{\cal G}(t,t')=0 \;\;\;\; \mbox{for} \;\; t<t'
\; .
\eeq
These equations do not contain derivatives with respect to
time; they have been neglected following the assumption
of slow variation of $\cl$ and $\gl$ for long times.

Evaluating eq. (\ref{geq}) in $t=t'$ implies either $\gl(t,t)=0$
(corresponding to the high temperature regime) or
\beq
2 (\tau-q) + 3 y q^2 = 0
\; .
\label{replicon}
\eeq
{}From here follows the value of the Edwards-Anderson parameter which is also
obtained in the static treatment \cite{Pa}.

Even if the eqs. (\ref{ceq}) and (\ref{geq}) are non-local,
they can be interpreted as asymptotic
if the crucial
assumption of weakness of the long-term memory is made:
eq. (\ref{wltm}) implies that the lower limit $t''=0$ in integrals such as
\beq
\int_{0}^{t} \; dt'' \; {\cal C}(t,t'') {\cal G}(t',t'')
\eeq
can be substituted by any lower limit
$t''=t_o$,
and this has no effect in the integral as long as $t$ and $t'$
both go to infinity.

As has been often noted the asymptotic equations
for $\cl$ and $\gl$
have an infinite set of
invariances \cite{So},\cite{Gi},\cite{Dofeio}.
Indeed, if we perform an arbitrary reparametrization of time:
\beq
{\hat t}=h(t) \; , \;\;\;\;\;\;\; {\hat t'}=h(t')
\label{reparam}
\eeq
with $h$ an increasing function, and we redefine
\beq
{\hat C}({\hat t},{\hat t'})=C(h(t),h(t')) \; , \;\;\;\;\;\;\;
{\hat G}({\hat t},{\hat t'})= h'(t') \, G(h(t),h(t'))
\; ,
\label{reparam1}
\eeq
the transformed
functions ${\hat C},{\hat G}$ satisfy the same equations in terms of the
\linebreak reparametrized times.
This means that given one solution we can obtain infinitely many others
by reparametrizations.

This invariance is a consequence of having neglected the time
derivatives in making the asymptotic limit. The full
dynamical equations have no such invariances; because of causality
their solution is {\em unique}.
The best we can do with eqs. (\ref{ceq}) and (\ref{geq}) is to find a
family of asymptotic solutions (related by reparametrizations). Which one
is actually the correct (unique) asymptote can only be decided from equations
that do not neglect time derivatives.
Throughout this work we shall try to go
as far as possible keeping the discussion at the reparametrization invariant
level: we shall only obtain solutions {\em modulo reparametrizations}
(however, the relaxation within a trap, as solved in Ref. \cite{Sozi}
is well determined and not affected by this invariance).

Having this in mind we shall make the following two further assumptions:

\vspace{1cm}

{\it iv.}
${\cal C}$ and ${\cal G}$ are related by a reparametrization
invariant formula. This can only be fulfilled by
\beq
{\cal G}(t,t')= X[{\cal C}(t,t')] \,
\frac{\partial {\cal C}(t,t')}{\partial t'} \; \theta(t-t')
\label{xeq}
\eeq
where $X$ depends on the times {\em only through} ${\cal C}$.
Indeed, under reparametrizations (\ref{reparam}) and (\ref{reparam1}) this
equation transforms into
\beq
{\hat {\cal G}}({\hat t},{\hat t}')= X[{\hat {\cal C}}
({\hat t},{\hat t}')] \;
\frac{\partial {\hat {\cal C}}({\hat t},{\hat t}')}{\partial {\hat t}'} \;
\; \theta({\hat t}-{\hat t}') \,
\eeq
{\it i.e.} it retains the same form.

Furthermore, if we supplement the definition of $X[z]$
with $X[z]=1$ for $q<z<1$
then the relation (\ref{xeq}) holds for all $C(t,t')$, $t'$ large ({\it cf}
eq. (\ref{fdt})).
$X[z]$ may be discontinuous in $z=q$, where it
jumps from $X[q]$ to $1$.

An immediate consequence of eq. (\ref{xeq})
is that all generalized susceptibilities (\ref{int})
are given by
\beq
I^r(t)= (1-q^r) + \int_{0}^{q} X[q'] \; d(q^{'r})
\; .
\label{aaa}
\eeq
Hence the dynamical
generating function of the generalized suceptibilities $P_d(q)$
(eq. (\ref{aa}))
and in particular the asymptotic energy
are entirely determined by the function $X[z]$ and $q$.

\vspace{1cm}

{\it v.} Given three large times $t_{min} \leq t_{int} \leq t_{max}$, the
corresponding three configurations
${\bf s}(t_{max})$, ${\bf s}(t_{int})$, ${\bf s}(t_{min})$
define a triangle the sides of which are given by
the three correlations
$C(t_{int},t_{min})$,
$C(t_{max},t_{min})$,
$C(t_{max},t_{int})$.
For the three times tending to infinity we propose
that the correlation between the extreme times
$C(t_{max},t_{min})$
is completely determined by
the correlations between the extreme times and any intermediate time:
\beq
C(t_{max},t_{min})= f[C(t_{max},t_{int}),C(t_{int},t_{min})]
\; .
\label{Feq}
\eeq
In other words, we are assuming that when $t_{min} \rightarrow \infty$
the correlation $C(t_{max},t_{min})$ only depends on $t_{min}$ and $t_{max}$
through
the other two correlations.
The relation $f$ is not necessarily smooth and we shall see
in the following sections that in fact it is not.
We can formally invert this equation by defining
the inverse function $\f$
\footnote{Note that in the definition of $\f$ the smallest argument is
always on the right}
\beq
C(t_{int},t_{min})= {\bar f}[C(t_{max},t_{int}),C(t_{max},t_{min})]
\; .
\label{fbeq}
\eeq
We are proposing relation (\ref{Feq})
for the whole range of values of $C$, including the FDT sector.

\vspace{1cm}

Both assumptions made in this section are amenable to numerical checks
and we shall discuss them in detail in Section \ref{VII}.

\vspace{.5cm}

\section{Triangle Relations}
\label{IV}
\renewcommand{\theequation}{\thesubsection.\arabic{equation}}

\vspace{.5cm}

In this section we shall study the properties of the function $f$,
defined in eq. (\ref{Feq}). {\em A priori} $f$
has no reason to be smooth, though we shall assume throughout that it
is continuous. All the results we shall present in this section
are general, they do not depend
on any particular model but just follow from assumption {\it v.} ({\it cf.}
eq. (\ref{Feq})).

\subsection{Basic Properties}
\setcounter{equation}{0}

The first trivial properties are
\beq
f(x,1)=f(1,x)=x
\label{neutr}
\eeq
which are obtained choosing $t_{int}=t_{min}$ and $t_{int}=t_{max}$,
respectively.

Since we are assuming that the system drifts away at any time,
namely assumption {\it i.},
\beqa
\begin{array}{rclcrcl}
 f(a,y) &\geq& f(b,y) \;\;\;\;\;\;\;\; &\mbox{if}& \; a& \geq &b
\; ,
\\
 f(y,a) &\geq& f(y,b) \;\;\;\;\;\;\;\; &\mbox{if}&  \; a& \geq &b
\; .
\end{array}
\label{desigual}
\eeqa
In particular for any $(a,b)$,
\beqa
\begin{array}{rcccl}
 f(a,1) &=& a &\geq & f(a,b)
\; ,
\\
 f(1,b) &=& b &\geq & f(a,b)
\; ,
\end{array}
\eeqa
and this implies
\beq
f(a,b) \leq \min (a,b)
\; .
\label{desimin}
\eeq

Consider now four successive times $t_1 <t_2 <t_3 < t_4$. We have
\beqa
C(t_{4},t_{1})&=& f[C(t_{4},t_{3}),C(t_{3},t_{1})] \nn
&=& f[C(t_{4},t_{2}),C(t_{2},t_{1})] \nn
&=& f[C(t_{4},t_{3}),f[C(t_{3},t_{2}),C(t_2,t_1)]] \nn
&=& f[f[C(t_{4},t_{3}),C(t_{3},t_{2})],C(t_2,t_1)]]
\; ,
\label{assoc}
\eeqa
{\it i.e.} $f$ is {\em associative}.

The existence of a neutral (eq.(\ref{neutr})) and the
requirement of associativity
severely restricts the choice of the function $f$.

Two important examples of functions (the first one not smooth, the
second one smooth) satisfying eqs. (\ref{neutr}),
(\ref{desigual}) and (\ref{assoc}) are:
\beq
f(a,b)= \min(a,b)
\; ,
\eeq
the ultrametric relation,
and
\beq
f(a,b)=ab
\; .
\label{rig}
\eeq

One can check that this
last relation corresponds to the vector ${\bf s}$ evolving in such
a way that the direction of the trajectory at two times is uncorrelated:
{\it i.e.} which spin flips at a given time is independent of which ones
flipped
before. The spherical triangle determined by ${\bf s}(t_{min})$,
${\bf s}(t_{int})$ and ${\bf s}(t_{max})$ is then, for probabilistic reasons,
right-angled.

As an example of the physical meaning of the function $f$, consider
the
slight variation of eq. (\ref{rig}) which
was found in \cite{Cuku} for the long-time
correlations $\cl$ of the $p$-spin spherical model
\beq
\frac {{\cl}(t_{max},t_{min})}{q}=\frac{{\cl}(t_{int},t_{min})}{q}
\frac{{\cl}(t_{max},t_{int})}{q}
\; .
\label{pspin}
\eeq
This can be understood if one defines the `magnetization' vector
for large times $t$ as
\beq
m_i(t) = \frac{1}{\Delta} \int_0^\Delta d{\cal T} s_i(t+{\cal T})
\eeq
($\Delta \rightarrow\infty$, $\Delta/t \rightarrow 0$). Then a short
computation shows that $(1/N) \sum_i m_i^2 \rightarrow q$ and
$\cl(t,t')/q$ is the cosine of the angle subtended by ${\bf m}(t)$ and
${\bf m}(t')$ at two widely separated times ($t-t' >> \Delta$). Then
eq. (\ref{pspin}) says that
the magnetization vector ${\bf m}(t)$
describes a trajectory without memory of the direction,
and hence makes right angled triangles in any three large
(and widely separated) times.

\subsection{Correlation scales}
\setcounter{equation}{0}

Consider now the function $f(a,a)$ satisfying ({\it cf.} eqs. (\ref{desigual}))
\beq
f(a,a) \leq a
\; .
\label{desi1}
\eeq
The above inequality admits fixed points $a^*_k$ such that
$f(a^*_k,a^*_k)=a^*_k$. These points can be isolated or they can form a
dense set. In fig. 1 we sketch a possible function $f(a,a)$.

Take now a succession of (large) times  $t_0<t_1<...<t_r$
such that the correlation between
two succesive times is $C(t_{i+1},t_{i})=b$,
 and compute the correlation $C(t_r,t_0)$ between
the two extremes of the succession
\beq
b^{(r)} \equiv C(t_r,t_0) = f(b,...f(b,f(b,b))...)
\; .
\eeq
The function $f$ is iterated $(r)$ times and the order of parenthesis
is immaterial because of associativity.
Choose two correlation values corresponding to
consecutive fixed points, say $a^*_1>a^*_2$ (see fig. 1), with
no other fixed points in between.
Then, it is easy to see that given two values $b_1,b_2$ with
$a^*_1>b_1>b_2>a^*_2$, there exists a  finite number  $(s)$ such that
\beq
b_1^{(s)} \leq b_2
\; ,
\label{rela}
\eeq
{\it i.e.} with a finite number of steps (iterations of $f$) with correlation
$b_1$ we can go up to or beyond $b_2$.
In contrast, if we consider a succession of steps each of correlation $a^*$
($a^*$ any fixed point), the correlation between the extremes never goes
beyond $a^*$ for finite $(s)$.

This suggests that we define {\em in a reparametrization-invariant way}
a correlation scale as the set of correlations that can be connected by
 relation (\ref{rela}) for some finite $(s)$.
This breaks the whole interval of correlations into equivalence classes.

With this definition a fixed point $a^*$ constitutes a
correlation scale in itself.
An interval which is made of
a dense set of fixed points such as  $(a^*_5,a^*_4)$ in fig. 1 is then a
dense set of scales.

The interval of correlations (none of them a fixed point of $f(a,a)$)
contained between two fixed points is a correlation scale which we
shall call `discrete'.

Note that the time difference ($t_{i+1}-t_{i}$)
needed to achieve a certain correlation is not
independent of the time $t_i$, but we  can suppose that it
does not decrease with time, {\it i.e.}
\beq
C(t_{i+1},t_i) = C(t_i,t_{i-1}) \;\; \Rightarrow \;\;
t_{i+1}-t_i \geq t_i-t_{i-1}
\; .
\eeq
Then the above definition of correlation scales
translates into a definition of `infinite
time-scales'.

\subsection{Ultrametric relations}
\setcounter{equation}{0}

Let us start by showing that the relation between fixed points is ultrametric.
Let $a^*_1$, $a^*_2$ be any two fixed points with $a^*_1>a^*_2$
\beqa
\begin{array}{rclcccl}
a^*_2
&=&
f(a^*_2,a^*_2)
&\leq&
f(a^*_2,a^*_1)
&\leq&
a^*_2
\; ,
\nn
a^*_2
&=&
f(a^*_2,a^*_2)
&\leq&
f(a^*_1,a^*_2)
&\leq&
a^*_2
\; ,
\end{array}
\eeqa
we have here  used eqs.
(\ref{desigual}) and (\ref{desimin}). Hence, for any two fixed points
\beq
f(a^*_1,a^*_2)=f(a^*_2,a^*_1)=\min(a^*_1,a^*_2)
\eeq
and ultrametricity holds.

Next, we consider a discrete scale limited by $a^*_i$ and $a^*_{i+1}$, two
consecutive fixed points, and a number $b$ such that $a^*_i>b>a^*_{i+1}$.
We assume that $f$ is a smooth function of $x,y$  within the scale.
Then, one has (see Appendix A):
\beqa
\begin{array}{rclcll}
f(a^*_i,b)
&=&
f(b,a^*_{i})
&=&
b
\;\;\;\;  &
\forall \; b \in (a^*_{i+1},a^*_i)
\; ,
\nn
f(a^*_{i+1},b)
&=&
f(b,a^*_{i+1})
&=&
a^*_{i+1}
\;\;\;\;  &
\forall \; b \in (a^*_{i+1},a^*_i)
\; ,
\end{array}
\eeqa
{\it i.e.} $a^*_{i}$ acts as the {\em neutral} and $a^*_{i+1}$ as the `zero'
inside the scale.

Consider now two different scales limited respectively by
$a^*_i>a^*_{i+1}$ and $a^*_k>a^*_{k+1}$.
Using the previous result we now
show that the relation $f(b_1,b_2)$ between two correlations
belonging to each scale
$a^*_i \geq b_1 \geq a^*_{i+1}$ and $a^*_k \geq b_2 \geq a^*_{k+1}$
is ultrametric:
\beq
f(b_1,b_2)=f(b_2,b_1)=\min(b_1,b_2)
\; .
\label{bbb}
\eeq
Let us start by choosing $b_1 = a_{i+1}^*$
and $a_{i+1}^* > a_k^*$  then
\beq
f(f(b_2,a^*_k), a^*_{i+1})
=
f(b_2, a^*_{i+1}) = f(b_2,f(a^*_k, a^*_{i+1}))
=
f(b_2, a^*_{k}) = b_2
\; .
\eeq
Hence
\beq
f(b_2, a^*_{i+1}) = b_2
\; ,
\eeq
and similarly
\beq
f(a^*_{i+1},b_2) = b_2
\; .
\eeq
Now choose $b_1$ satisfying $a_{i+1}^* < b_1 \leq a_i^*$, then
\beq
f(b_2,b_1)
=
f(f(b_2,a^*_k), b_1) = f(b_2, f(a^*_k, b_1))
=
f(b_2,a^*_k) = b_2
\; .
\eeq
Hence
\beq
f(b_2,b_1) = b_2
\; ,
\eeq
and similarly
\beq
f(b_1,b_2) = b_2
\; .
\eeq

Within a scale ultrametricity does not hold but
if we assume that
$f$ is there smooth it is then a one dimensional formal group law and
we have
for $b_1, b_2$ within the $k$th-scale that $f(b_1,b_2)$ is of the form
\beq
f(b_1,b_2) = \jmath^{-1}_k [ \jmath_k(b_1) \cdot \jmath_k(b_2) ]
\eeq
for some function $\jmath_k(x)$ which can be
different for each scale.
This is a well-known result from
formal group theory \cite{Ha} and we present it in Appendix B.

Furthermore, this implies that within a scale (see eq. (\ref{domi}))
\beq
C(t,t') = \jmath^{-1}_k \left( \frac{h_k(t')}{h_k(t))} \right)
\label{ngo}
\eeq
for some increasing function $h_k(x)$ ( Appendix B).
The solution in Ref. \cite{Cuku} corresponds to only one
discrete scale (apart from the FDT scale), and has the form of eq. (\ref{ngo}).

\vspace{.5cm}

\section{Dynamical equations}
\label{V}
\setcounter{equation}{0}
\renewcommand{\theequation}{\thesection.\arabic{equation}}

\vspace{.5cm}

In this section we shall study the asymptotic dynamical equations
of the SK model (eqs. (\ref{ceq})
and (\ref{geq})). We shall start by simplifying the equations using
assumption {\it iv.}

Let us define two functionals
\beqa
F[\cl] &=& - \int_\cl^q \; d\cl' \; X[\cl']
\; , \\
H[\cl] &=& - \int_\cl^q \; d\cl' \; {\cl'}^2 X[\cl']
\; .
\eeqa
Note that even if $X$ may be discontinuous, $F$ and $H$ are continuous.
Thus inserting eq. (\ref{xeq}) in
eqs. (\ref{ceq}) and (\ref{geq}) we get
\beqa
-
3 y q^2 \cl(t,t')
+
y \, \cl^3(t,t')
+
\int_{0}^{t'} \; dt'' \;
\cl(t,t'')
\frac{\partial F[\cl(t',t'')]}{\partial t''}
\;\;\;\;\;\;\;\;\;\;\;\;\;\;\;\;\;\;\;\;
& &
\nn
+
\int_{0}^{t'} \; dt'' \;
\frac{\partial F[\cl(t,t'')]}{\partial t''}
\cl(t',t'')
+
\int_{t'}^{t} \; dt'' \;
\frac{\partial F[\cl(t,t'')]}{\partial t''}
\cl(t'',t')
&=&
0
\nn
& &
\label{ceq1}
\\
-3 y q^2
\frac{\partial F[\cl(t,t')]}{\partial t'}
+ \;
\int_{t'}^{t} \; dt'' \;
\frac{\partial F[\cl(t,t'')]}{\partial t''}
\frac{\partial F[\cl(t'',t')]}{\partial t'}
+
3y \,
\frac{\partial H[\cl(t,t')]}{\partial t'}
&=&
0
\nn
\label{geq1}
\eeqa
This last equation can be integrated once w.r.t. $t'$ to give
\beq
-3 y q^2 F[\cl(t,t')]
+ \;
\int_{t'}^{t} \; dt'' \;
\frac{\partial F[\cl(t,t'')]}{\partial t''} F[\cl(t'',t')]
+
3y \,
H[\cl(t,t')]
=
0
\; .
\label{geq2}
\eeq
We now have two reparametrization invariant equations
for $\cl(t,t')$ which have to be satisfied simultaneously.
Eqs. (\ref{ceq1}) and (\ref{geq2})
can be written in a form in which the times disappear
using assumption {\it v.} Indeed, with definition (\ref{fbeq}) they
formally become
\beqa
-3 y q^2 \cl +
\int_{0}^{\cl} \; d\cl' \;
X[\cl'] \f(\cl,\cl')
-
\int_{0}^{\cl} \; d\cl' \;
F[\f(\cl,\cl')]
& &
\;\;\;\;\;\;\;\;\;\;\;
\nn
+
\int_{\cl}^{q} \; d\cl' \;
X[\cl'] \f(\cl',\cl)
+ y \, \cl^3
&=&
0
\; ,
\label{ceq3}
\\
-3 y q^2 F[\cl]
+ \;
\int_{\cl}^{q} \; d\cl' \;
X[\cl'] F[\f(\cl',\cl)]
+
3y \,
H[\cl]
&=&
0
\; ,
\label{geq3}
\eeqa
since $\f(\cl,\cl')$ is
not necessarily a well-behaved function on the whole interval
$[0,1]$ we shall take care of this point in the next subsection.
Eqs. (\ref{ceq3}) and (\ref{geq3}) together determine
$X[\cl]$ and $\f(\cl,\cl')$.
Having eliminated the explicit time dependence,
we have effectively divided by
the reparametrization group. In what follows we shall concentrate
on studying the solution to these equations.

\subsection{Equations within a discrete scale}
\setcounter{equation}{0}
\renewcommand{\theequation}{\thesubsection.\arabic{equation}}

Let us take $\cl$ belonging to a discrete scale, $\cl \in (a_2^*,a_1^*)$.
Due to the ultrametricity between different scales ({\it cf.} eq.
(\ref{bbb})) we can use
\beq
\f(\cl,\cl') = \cl'
\; ,
\eeq
when $\cl$ is outside the scale of $\cl'$. This allows to simplify
eqs. (\ref{ceq3}) and (\ref{geq3}) and furthermore to deal with
a region where the function $\f(\cl,\cl')$ is smooth.
In these regions  $\f(\cl,\cl')$ can be written as (see Appendix B)
\beq
\f(\cl,\cl') = \jmath^{-1} \left( \frac{\jmath(\cl')}{\jmath(\cl)} \right)
\;\;\;\;\;\;\;\;\;\; \mbox{for} \; \cl \geq \cl'
\label{fj}
\eeq
and since $\jmath(a_1^*)=1$
\beq
\f(a_1^*,\cl) = \cl
\; .
\eeq
Therefore the equations become
\beqa
-3 y q^2 \cl +
a_2^* F[a_2^*] - \cl F[a_1^*] + y \cl^3
-
2
\int_{0}^{a_2^*} \; d\cl' \;
F[\cl']
\;\;\;\;\;\;\;\;\;\;\;\;\;\;\;\;\;\;\;
& &
\nn
+
\int_{a_2^*}^\cl \; d\cl' \;
X[\cl'] \f(\cl,\cl')
-
\int_{a_2^*}^\cl \; d\cl' \;
F[\f(\cl,\cl')]
+
\int_{\cl}^{a_1^*} \; d\cl' \;
X[\cl'] \f(\cl',\cl)
&=&
0
\nn
& &
\label{ceq4}
\\
-3 y q^2 F[\cl]
+
3y \,
H[\cl]
-
F[\cl] F[a_1^*]
+
\int_{\cl}^{a_1^*} \; d\cl' \;
X[\cl'] F[\f(\cl',\cl)]
&=&
0
\nn
\label{geq4}
\eeqa

Evaluating eq. (\ref{geq4}) in $\cl = a_1^*$ and $\cl = a_2^*$ respectively
we have
\beqa
3 y H[a_1^*] &=& F[a_1^*] ( F[a_1^*] + 3 y q^2)
\; ,
\label{h1}
\\
3 y H[a_2^*] &=& F[a_2^*] ( F[a_2^*] + 3 y q^2)
\; .
\label{h2}
\eeqa
Moreover differentiating eq. (\ref{geq4}) w.r.t. $\cl$ and evaluating in
$\cl=a_1^*$ we obtain
\beq
X[a_1^*] ( -3 y q^2 - 2 F[a_1^*] + 3 y {a_1^*}^2 ) = 0
\; ,
\eeq
and this implies $X[a_1^*] = 0$ or
\beq
F[a_1^*] = \frac{3y}{2} ({a_1^*}^2-q^2)
\; .
\label{F1}
\eeq

Differenting twice eq. (\ref{geq4}) w.r.t. $\cl$ and evaluating in
$\cl=a_1^*$ we have
\beq
X[a_1^*] \left( 6 y a_1^*  - X[a_1^*]
\left.
\frac{\partial \f(\cl',\cl)}{\partial \cl} \right|_{\cl'=\cl=a_1^*}
\right)
=
0
\; .
\eeq
The derivative on the right can be calculated using eq. (\ref{fj})
and is equal to one. Then if
$X[a_1^*] \neq 0$
\beq
X[a_1^*] = 6 y a_1^*
\; .
\label{equits}
\eeq

\subsection{X within a discrete scale}
\setcounter{equation}{0}

We shall assume that

{\it vi.} $X$ is a non decreasing function.

\noindent Under this last assumption, we shall show that $X$ is
constant within a discrete scale \cite{Crhoso}.

Differentiating eqs. (\ref{ceq4}) and (\ref{geq4}) w.r.t. $\cl$, multiplying
the first one by $X[\cl]$ and subtracting, we obtain
\beqa
\int_{a^*_2}^\cl
d \cl' X[\cl]
\frac{\partial \f(\cl,\cl')}{\partial \cl}
\left(
X[\f(\cl,\cl')] - X[\cl']
\right)
\;\;\;\;\;\;\;\;\;\;\;\;\;\;\;\;\;\;\;\;
& &
\nn
+
\int_\cl^{a^*_1} d\cl' \;
X[\cl']
\frac{\partial \f(\cl',\cl)}{\partial \cl}
\left(
X[\f(\cl',\cl)] - X[\cl]
\right)
&=&
0
\; .
\label{dg-dc}
\eeqa
Since this equation is valid within any scale, we see that a sufficient
condition for eqs. (\ref{ceq4}) and (\ref{geq4}) to be compatible is that
$X$ be constant within each scale. Let us now show that if $X$ is
non-decreasing this is the only possibility. Differentiating
eq. (\ref{dg-dc}) w.r.t. $\cl$ and then evaluating in $\cl = a^*_1$
yields
\beq
X[a^*_1] \int_{a^*_2}^{a^*_1} \; d\cl'
\left.
\left(
\frac{\partial \f(\cl,\cl')}{\partial \cl}
\right)^2
\right|_{\cl=a^*_1} X'[y] \;\;\; = \;\;0
\; .
\eeq
This equation admits the solution $X[a^*_1] =0$ which corresponds to the
high temperature phase. If this is not the case,
and assuming that $X'[z] \geq 0$, the integrand should vanish.
The squared factor looks like
\beq
\left.
\frac{\partial \f(\cl,\cl')}{\partial \cl}
\right|_{\cl=a_1^*}
=
\jmath'(a_1^*) \cdot \frac{\jmath(\cl')}{\jmath'(\cl')}
=
\left.
\frac{\partial \ell(z)}{\partial z}\right|_{z=a_1^*} \;
\left(
\frac{\partial \ell(\cl')}{\partial \cl'}
\right)^{-1}
\eeq
(see Appendix B) and does not vanish for $\cl' > a_2^*$. Thus,
\beq
X'[y] = 0 \;\;\;\;\;\;\;\;
\Rightarrow  \;\;\;\;\;\;\;\;
X[y]  =  X \;\;\;\;  \mbox{constant}
\eeq
for $y \; \in (a_2^*,a_1^*)$.
Hence, using eq. (\ref{equits})
\beq
X[y] = X = 6 y a_1^*
\; .
\eeq

\subsection{No discrete scales}
\setcounter{equation}{0}

Let us show that the discrete scales collapse.
Using that $X$ is constant within the discrete scale,
we have
\beqa
F[a_2^*] &=& F[a_1^*] + X (a_2^* - a_1^*)
\; ,
\label{f3}
\\
H[a_2^*]  &=& H[a_1^*] + \frac{X}{3} ({a_2^*}^3 - {a_1^*}^3)
\; .
\label{h3}
\eeqa
Putting this into eqs. (\ref{h1}) and (\ref{h2}) and
using eq. (\ref{F1})
\beq
(a_2^* - a_1^*)^2 [y (a_2^* + 2 a_1^*) - X ] = 0
\; .
\eeq
This equation gives as one possible root
$a_2^* = 4 a_1^*$,  which is  not acceptable since
by hypothesis $a_2^* \leq a_1^*$. Hence we are left with
\beq
a_2^* = a_1^*
\eeq
and this implies that each discrete scale is indeed empty.

Having shown that there are no discrete scales (except for the FDT scale,
which is not contained in the previous equations), one sees
that the solution should verify ultrametricity for all values
of the correlations $\cl$.

\vspace{.5cm}

\section{Ultrametric Solution}
\label{VI}
\setcounter{equation}{0}
\renewcommand{\theequation}{\thesection.\arabic{equation}}

\vspace{.5cm}

Let us now describe in detail the ultrametric solution:
for $t_{min} \rightarrow\infty$,
\beqa
\begin{array}{rcl}
C(t_{max},t_{min})
=
\min (C(t_{max},t_{int}), C(t_{int},t_{min}))
\;
&
\mbox{if}
&
\; C(t_{max},t_{min}) \leq q
\end{array}
\label{ultx}
\eeqa
and if $C(t_{max},t_{min}) \geq q$ the solution of Sompolinsky-Zippelius
holds.

In the preceding
section we have concluded that we have a dense set of scales,
so that
\beq
\f(\cl,\cl') = \cl'
\eeq
$\forall \cl' < \cl$. Thus eqs. (\ref{ceq3}) and (\ref{geq3}) simplify to
\beqa
- 3 y q^2 \cl - \cl F[\cl] + y \cl^3
+ \int_0^\cl d\cl' \, \cl' \, X[\cl']
- \int_0^\cl d\cl' \, F[\cl']
&=&
0
\; ,
\label{ceq31}
\\
- 3 y q^2 F[\cl] - (F[\cl])^2 + 3 y H[\cl]
&=&
0
\; ,
\label{geq31}
\eeqa
and from here follows that
eqs. (\ref{h1}) and (\ref{F1}) hold for every value $\cl<q$:
\beqa
3 y H[\cl] &=& F[\cl] (F[\cl] + 3yq^2)
\; ,
\nn
F[\cl] &=& \frac{3y}{2} (\cl^2-q^2)
\; .
\label{FF1}
\eeqa
Differentiating eq. (\ref{FF1}) w.r.t. $\cl$ we obtain:
\beq
X(\cl)=3 y \cl
\; .
\eeq

This yields for $P_d(q)$ ({\it cf.} eqs. (\ref{int}), (\ref{aa}) and
(\ref{aaa}))
\beq
P_d(q')=(1-X[q']) \; \delta(q'-q)+ 3 y U(q')
\; ,
\eeq
where $U(q')=1$ if $0<q'<q$, and zero otherwise.
The value of $q$ is given by eq. (\ref{replicon}).
Hence, we have found that
\beq
P_d(q')=P(q')
\eeq
for the SK model where $P(q')$ is the Parisi functional
order parameter associated with the Gibbs-Boltzmann measure \cite{Pa},
also implying that the dynamical and statical transition temperatures
coincide.
This equality is not obvious,
and is a property of this particular model. Indeed, this same
dynamics yields for the model of Ref. \cite{Cuku} a dynamical $P_d(q)$ which
is {\em different} from the static one.
For the SK model
the energy  and susceptibility {\em to leading order in $N$}
coincide with the
corresponding equilibrium values,
and the size of the `traps' encountered for large
times coincides with the size of the equilibrium states.

In particular, ultrametricity implies that a plot of $C(t,t')$ {\it vs.}
$t'$ tends to have  a long plateau. More precisely, consider the function
\beq
\hat C(\mu) = \lim_{t\rightarrow\infty} C(t,\mu t)
\; .
\eeq
It is easy to see that eq. (\ref{ultx}) implies that $\hat C(\mu)$
drops from one to a certain value $\tilde q$ ($0 \leq \tilde q \leq q$)
in a small neighbourhood of $\mu = 1$, remains constant and equal
to $\tilde q$ in the interval $(0,1)$ and drops from $\tilde q$ to
zero in a small neighbourhood of $\mu = 0$. The actual value of
$\tilde q$ cannot be determined unless one goes beyond the reparametrization
invariant results. This last result was verified in ref. \cite{Frme}
for the model studied there.

\vspace{.5cm}

\section{Simulations and Measurable Results}
\label{VII}
\setcounter{equation}{0}

\vspace{.5cm}

In this section we discuss some consequences of the assumptions and derivations
of the previous sections that can be checked with numerical simulations.
Some of these involve magnitudes that are measurable experimentally;
for a toy model such as the SK this is not such an advantage, but it would
be if some of these results turn out to hold also for finite-dimensional
systems.

Our results have been obtained in the thermodynamic limit for asymptotic
times,  $\lim_{t\rightarrow\infty}$ after $\lim_{N\rightarrow\infty}$.
In a simulation this means that one has to
eliminate finite-size effects.

Firstly, the two initial assumptions {\it i.} and {\it ii.} have been
already observed in Ref. \cite{Cukuri}
where the out of equilibrium dynamics of the hypercube spin glass
has been studied numerically. This model is expected to reproduce
for high dimensionalities the SK model \cite{Pariru,Gemeye}.

Assumption {\it i.}, weak ergodicity breaking, has been verified
by plotting the correlation function $C(t+t_w,t_w)$ {\it vs.} $t$ in a
log-log plot for different waiting times $t_w$ (Fig. 3 in Ref.
\cite{Cukuri}).
Furthermore, numerical simulations of the $3D$ Edwards-Anderson
model also support this assumption in that realistic model \cite{Ri}.

Assumption {\it ii.}, weakness of the long-term memory, or
equivalently the decaying to zero of the thermoremanent magnetization,
has also been verified
(Fig. 2 of Ref. \cite{Cukuri}). A similar
behaviour has been obtained both for realistic models \cite{Anmasv}
and experimentally \cite{Lusvnobe,Vihaoc}.

Secondly,
as a consequence of eq. (\ref{xeq})
the response $\chi (t,t_w)$ introduced in eq. (\ref{wltm}) is given by
\beq
\chi (t,t_w)= \int_0^{C(t,t_w)} \; dq' \; X[q'] = F[C(t,t_w)] - F[0]
\equiv \tilde F[C(t,t_w)]
\; .
\label{univers}
\eeq
Hence,
for large enough times $t$ and $t_w$, the times $(t,t_w)$ enter parametrically
in a plot $\chi (t,t_w)$ {\it vs.} $C(t,t_w)$
and all the points obtained for different pairs $(t,t_w)$
should lie on a {\em single} universal curve,
the integral function $\tilde F$ of $X$.

 The plot
$\chi (t,t_w)$ {\it vs.} $C(t,t_w)$ for the hypercubic cell of dimension
$D=15$ at temperature $T=0.2$
is shown in Fig. 2, together with the second integral of the
static $P(q')$ evaluated in $C(t,t_w)$ for the SK model.
The curves for different $t_w$ roughly coincide;
the departure is not systematic w.r.t. $t_w$. However, they do not coincide
with the corresponding static curve for the SK model at that temperature.
This could be an effect
of the finite dimension $D$, not inconsistent with the static results of
ref. \cite{Pariru}. It was here found
that the function $P(q')$ for small $q'$ is
is quite smaller for the hypercubic cell of dimension $D=12$
than for the SK model.

Thirdly, assumption {\it iv.} on the existence of the triangle relations
and its implicancies
can be tested numerically in the following way: \n
a) choose a number $\cal C$ and a large number $t_w$.\n
b) determine the time $t_{max}$ such that $C(t_{max},t_w)$=${\cal C}$.
\n
c) plot for all times $t$ ($t_w<t<t_{max}$) $C(t_{max},t)$ versus
$C(t,t_w)$. \n
d) repeat the procedure for a larger $t_w$ and the same ${\cal C}$. \n
The limiting curve, obtained as $t_w$ becomes
larger, is the (implicit) function given by eq. (\ref{Feq})
\beq
{\cal C}=f(x,y)
\; ,
\eeq
or
\beq
y = \f(x,\cl)
\; .
\eeq

If, as has been found in the previous sections, ultrametricity holds,
then the area limited by the horizontal line $x\in[\cl,1]$, $y=\cl$,
the vertical line $x=\cl$, $y\in [\cl,1]$ and the curve constructed
following the procedure above, would vanish when $t_w\rightarrow\infty$.
Studying the behaviour of this area is more practical than simply looking at
the curves since its calculation involves many points and
reduces the noise.
In Fig. 3 we present
the curves obtained in this way ({\it n.b} these curves have been
smoothed using a local interpolation in order to better show the
qualitative tendency, error bars of order $\simeq 0.01$ should be
taken into account).
In the inset we include a log-log plot of the area
{\it vs.} $t_w$.
The
approach of the curves to their limit is very slow and this
could raise the suspicion that this is a finite size effect.
We have checked however that for a system four times smaller the decrease in
area is not very different. The qualitative trend does not depend
on the temperature; we have also checked these results for higher subcritical
temperatures though we shall not present them here. Finally, note that
the value of $q$ ($\simeq 0.92$ for $T=0.2$) is easily seen in the
Figure.

Because the correlation functions, even though easy to compute numerically,
are hard to measure in an experimental system, it is convenient
to have a relation involving only the easily measurable quantity $\chi$.
Consider the function ${\cal F}$ defined by
\beq
\chi(t,t')= {\cal F}( \chi(t,t''), \chi(t'',t'))
\; .
\eeq
{}From the preceding sections we have that
\beq
{\cal F}( x,y)={\tilde F} [ f ( {\tilde F}^{-1}[x],{\tilde F}^{-1}[y])]
\; ,
\eeq
so that if $f$ is associative (commutative) then ${\cal F}$ also is.
Indeed,
${\cal F}$ and $f$ are isomorphic as a group law \cite{Ha}, and all results
quoted for the function $f$ carry through to ${\cal F}$.
In the ultrametric case, since ${\tilde F}$ is a growing function, we have
\beq
{\cal F}(x,y)= \min (x,y)
\; .
\eeq
Although the results of the present paper are in principle only valid
for mean-field systems, we have not resisted the temptation to check this
last relation
with the experimental data for spin-glasses of \cite{Vihaoc}, with negative
results. The fuction $\chi(t,t')$ for experimental systems follows an almost
perfect $t'/t$ law, and is in this sense rather more similar to a system
with discrete scales.

A detailed analysis of the numerical solution of the mean-field
dynamical equations corresponding to the relaxation of a particle
in a random potential in infinite dimensions was carried out in
Ref. \cite{Frme}. In particular it is verified there that
the results for large waiting times coincide to order $N$ with
great precision with those of the static treatment. This coicidence
is also found here for the SK model, but not in
the $p$-spin spherical model (Ref. \cite{Cuku}).
The common element of the former two models is that the replica analysis
of their statics involves an infinite number of breakings of the
replica symmetry, while the $p$-spin model has statically
only one breaking and dynamically one discrete scale apart from the
FDT scale.

\vspace{.5cm}

\section{Discussion}
\label{VIII}
\setcounter{equation}{0}

\vspace{.5cm}

We are now in a position to discuss qualitatively the out of
equilibrium dynamics of this model.

The system first relaxes rather rapidly to an energy which is slightly above
the equilibrium energy. In that region it starts encountering `traps'. Within
a trap the relaxation is rapid and described by the Sompolinsky-Zippelius
dynamics \cite{Sozi}.

As time passes, the energy relaxes slowly towards the equilibrium value,
the $O(N)$ difference between the dynamical and equilibrium energy goes
to zero.
The actual states contributing to equilibrium have an energy differing by
only $O(1)$ from the lowest state \cite{Mepavi}, and are not reached in finite
times. However, their cousins the long-time traps resemble them
in their geometry (value of $q$, relaxation within them), except that the
barriers surrounding a true state are divergent, while those surrounding
traps are finite, though large.
The evolution away from traps becomes slower and slower as time passes,
so that traps encountered at longer times tend to resemble more and more
the actual states contributing to equilibrium.

The fact that the system relaxes to values that are equal to order $N$ to
the equilibrium values (energy, $P_d(q)$, etc.)
and that the dynamical and static critical temperature coincide
{\em is not} a general feature of spin-glass out of equilibrium dynamics,
but a property
of this model. A different situation has been found for the $p$-spin
spherical model \cite{Cuku}, for which the non-equilibrium dynamics goes to
a threshold level that is above (to $O(N)$) the states that contribute to
the Gibbs-Boltzmann measure (and hence $P_d(q)\neq P(q)$ and the asymptotic
energy is different from the Gibbs-Boltzmann measure energy).
The reason for this difference can be seen by considering the TAP approach.
In the $p$-spin spherical model, all the TAP
valleys below the threshold
are separated by infinite barriers and have a positive-definite
Hessian. The system would remain trapped within any of them, but it never does
because it stays touring at the threshold energies above which there
are no valleys and below which the barriers are $O(N)$, {\it i.e.} in the
small range of free energies in which the barriers are $O(1)$.

In the SK model there seems to be no threshold of this kind, in the sense that
 the TAP valleys of the free energy encountered above the equilibrium free
energy should be separated by only $O(1)$ barriers. This is quite reasonable
if one accepts that most of these solutions (unlike the ones of the $p$-spin
spherical
model) are `born' by division of other solutions as the temperature is
lowered and, moreover, their Hessian contains zero eigenvalues \cite{Brmo}.

We also note that both the hypotheses
of weak ergodicity breaking and of weakness of the long term memory
can be understood within this scenario.
Since no traps are true states, the system eventually drifts away
forgetting the characteristics of any given trap (and in particular
its magnetization).

\vspace{.5cm}

\section{Conclusions}
\label{IX}
\setcounter{equation}{0}

\vspace{.5cm}

We have presented the relaxational dynamics of the
SK model in the thermodynamic limit in a way that naturally
involves an asymptotic out of equilibrium regime and aging effects.
We have restricted ourselves to a situation in which the system is
rapidly cooled to a sub-critical temperature and every external parameter
is afterwards left constant, as in some experimental settings.
A different approach to the dynamics is to consider explicitly changes
in the external parameters such as the temperature \cite{Dofeio,Ho}.

We have shown that the asymptotic dynamical equations can be solved
under mild assumptions which we have tried to state as explicitly as
possible. Since we are not allowing for the crossing of infinite
barriers all the assumptions can be checked numerically with relatively
small computer times and we
have presented preliminary results in this direction. We have
derived a set of equations containing only the correlation function
and the relation between three correlations (triangle relation)
and we have found that the unique solution to these equations
implies ultrametricity for every three widely separated times.
Without the assumption of time homogeneity and in the absence
of any parameter controlling the scales, it is not {\it a priori}
obvious how to define these scales.
We have given a precise definition of correlation scales, which
can be applied to other models.

The present treatment has important common elements with both the
static replica analysis and the dynamics \`a la Sompolinsky.

As regards the former this is due to an underlying formal
algebraic similarity between the replica treatment
and the asymptotic dynamics. Quite generally
the asymptotic dynamical equations can be derived using
this similarity and their solution has a connection with
the replica solution with an ansatz \`a la Parisi (although
this does not necessary mean that the statics and asymptotic
dynamics should give the same results
for every model) \cite{Cufrkume}.

It is worthwhile discussing in more detail the similarities and differences
with Sompolinsky's dynamics.
In that framework, one assumes time-homogeneity plus a relation
like (\ref{xeq}) between the correlation
and response functions.
Then one further assumes
that the correlation function relaxes (ever more slowly)
to zero for widely separated times
(in the absence of a magnetic field).
With these assumptions one hopes
to have a representation of the equilibrium dynamics after an
infinite waiting time and for large but finite $N$.

A well-known problem of this picture is that
the decaying to zero of the correlations is incompatible with the
equilibrium solution \cite{Mepavi}
unless one considers multiple dynamical
solutions for times long enough as to allow for infinite-barrier
crossing \cite{Frku}.
Moreover, there is an additional puzzle:
the hypothesis of time homogeneity applied to the
$p$-spin spherical model fails to give the equilibrium values
\cite{Crhoso}.

After the work of Sompolinsky, Ginzburg considered the effect of
a perturbation on a spin glass which is already in equilibrium
\cite{Gi}. This is different from our approach since we do not
assume that the system has reached equilibrium and we find
that it never does. Furthermore the mechanism for
barrier crossing we here invoke is not related to the size of the
system but rather to its age.

As mentioned in the introduction we work here with $N$ infinite
and any infinite time scale arises not because $N$ or any other
parameter go to infinity but it is the very age of the system
which imposes the rhythm of the relaxation, which eventually
becomes very slow. The assumption that the correlations go to
zero in this context is just related to weak ergodicity breaking
and the observable aging effects and does not contradict the statics.
Furthermore, since we are considering finite albeit long times
the solution for the correlation and response functions is unique.
Yet the fast relaxation for small time differences (but large overall
times) is identical
to the one found by Sompolinsky-Zippelius with the only change
of reinterpreting the state as a trap or channel from which the system
always escapes.

As noted in section \ref{IV}, all the results in this paper are
invariant with respect to reparametrizations in time. This invariance
is not a true property of the dynamics, but is the result of using equations
of motion that are, for the large time-differences, only asymptotically
valid.
 The true solution has to choose one asymptote between all the family of
reparametrizations we have obtained.
This problem is quite common in the asymptotic matching of the
solutions to differential equations. The application of such concepts to the
spin glass problem is still an open question.
Many of the most interesting results (decay law of the energy, of the
magnetization, etc) will only be available when one will be able to
go beyond reparametrization invariance.

There are quite a few open questions even at the level of reparametrization
invariant results. One would like to have a deeper understanding of the
function $X(C)$
and the dynamical generating function $P_d(q)$. It may be that some general
theorems can be derived for the  asymptotic non-equilibrium regime.

Some of the questions discussed here are  quite general, and it would be
interesting to try them in other  models.

\vspace{2cm}

{\bf Aknowledgments}
We wish to aknowledge useful discussions and suggestions from J-P. Bouchaud,
 S. Franz, E. Marinari, M. Mezard , G. Parisi, F. Rodriguez Villegas,
and M.A. Virasoro.

\vspace{2cm}

{\Large {\bf {Appendix A}}}
\setcounter{equation}{0}
\renewcommand{\theequation}{A.\arabic{equation}}

\vspace{1cm}

In this Appendix we show that assuming that $f(x,y)$ is smooth within
a discrete scale, the upper limit of the scale acts as
the {\it neutral} ({\it i.e.} like the identity in a product) and that the
lower limit acts as the {\it zero} in the product.

The associativity relation (\ref{assoc}) implies
\beq
\left.
\frac{\partial f(x, f(y,z))}{\partial x}
=
\frac{\partial f(w,z)}{\partial w} \right|_{w=f(x,y)}
\frac{\partial f(x,y)}{\partial x}
\; .
\eeq
Consider an isolated fixed point $a^*_1$ separating two discrete scales
and set $y=z=a^*_1$ in the above equation:
\beq
\frac{\partial f(x,a^*_1)}{\partial x}
\left(
\left.
1 -
\frac{\partial f(w,a^*_1)}{\partial w}
\right|_{w=f(x,a^*_1)}
\right)
= 0
\; .
\label{apA1}
\eeq
This admits the solution
\beq
f(x,a^*_1) = g(a^*_1)
\; .
\label{res1}
\eeq
Since $f(x,a_1^*)$ is continuous and $f(a_1^*,a_1^*) = a_1^*$,
$g(a_1^*) = a_1^*$.
For $x>a^*_1$ this is a possible solution. For $x<a^*_1$, $f(x,a^*_1) \leq
x < a^*_1$ so this solution in not admitted in this case.

The other solution to eq. (\ref{apA1}) is
\beq
\left.
\frac{\partial f(w,a^*_1)}{\partial w}
\right|_{w=f(x,a^*_1)} = 1
\; .
\label{apA2}
\eeq

Consider the successive fixed point $a_2^* < a^*_1$. $f(x,a^*_1)$ satisfies
\beq
a^*_2 \leq f(x,a^*_1) \leq a^*_1
\eeq
and, because of eq.(\ref{desigual}), $f(x,a^*_1)$ is a monotonically
increasing function of $x$.
Equality on the left takes place when
$x=a_2^*$ and
on the right when $x=a^*_1$.
Hence, for every $w$ within the discrete scale we can solve $w=f(x,a^*_1)$
for $x$. Eq. (\ref{apA2}) becomes
\beq
\frac{\partial f(y,a^*_1)}{\partial y} = 1
\eeq
$\forall y \in (a^*_2,a^*_1)$. The solution is $f(y,a_1^*)=y + k(a_1^*)$;
since $f(a_1^*,a_1^*)=a_1^*$ then $k(a_1^*) = 0$ and we finally have
\beq
f(y,a^*_1) = y
\; .
\label{res2}
\eeq
This solutions is only possible if $y < a^*_1$. Otherwise
$y=f(y,a^*_1)\leq a^*_1$ and this is incompatible with $y> a^*_1$.

Hence, for $x$ within a discrete scale, $a_2^* < x < a_1^*$,
eqs. (\ref{res1}) and (\ref{res2}) imply
\beqa
\begin{array}{rcl}
f(x,a_2^*) &=& a^*_2
\; ,
\\
f(x,a^*_1) &=&  x
\; ,
\end{array}
\label{extremos}
\eeqa
and, $a_2^*$ and $a_1^*$ are the {\it zero} and {\it neutral}, respectively.

\vspace{1cm}

{\Large {\bf {Appendix B}}}
\setcounter{equation}{0}
\renewcommand{\theequation}{B.\arabic{equation}}

\vspace{1cm}

In this Appendix we review some results of formal group theory that give
the general form of $f(x,y)$ {\em within a discrete scale}.
Let $a^*_1$  and $a^*_2$ be two consecutive fixed points   $a^*_1>a^*_2$.
We assume that $f(x,y)$ is smooth for $x,y \in (a^*_2,a^*_1]$.
{}From (\ref{extremos}) we have that $a^*_1$ is the neutral element within
this range.
Under this assumption one can show that \cite{Ha}: $f(x,y)$ is
commutative and can be written as
\beq
f(x,y)=\ell^{-1} \circ (\ell(x)+\ell(y))
\label{ell}
\eeq
with $\ell(z)$ given by
\beq
\ell(z)=-\int^{a^*_1}_{z} \; dz' \; \left( \left.
\frac{\partial f(w,z')}{\partial w} \right|_{w=a_1^*} \right)^{-1}
\; ,
\eeq
$\ell(a^*_1)=0$.
Because we are within a discrete
scale the denominator in the integral is positive
definite for $z' \in (a^*_2,a^*_1]$ and it first vanishes in $z'=a^*_2$.
The function $\ell(z)$ is increasing and negative semi-definite
$(\ell(a^*_1)=0$).

We can also define
\beq
\jmath(z)= \exp(\ell(z))
\eeq
to obtain
\beq
f(x,y)=\jmath^{-1} \circ (\jmath(x) \cdot \jmath(y))
\; .
\label{jmat}
\eeq

Writing (\ref{ell}) in terms of the correlations at three times
\beq
\ell \circ C(t_{max},t_{min})=\ell \circ C(t_{max},t_{int})
+
\ell \circ C(t_{int},t_{min})
\label{ellell}
\eeq
the crossed second derivative vanishes
\beq
\frac{ \partial^2 \; \ell \circ C (t_{max},t_{min})}{\partial t_{max}
\partial t_{min} }=0
\; .
\eeq
The solution to this equation is
\beq
\ell \circ C(t_{1},t_{2})= {\tilde h}_1(t_1)-{\tilde h}_2(t_2)
\eeq
for some functions ${\tilde h}_1$, ${\tilde h}_2$. Inserting this into
(\ref{ellell}) we see that ${\tilde h}_1(t)={\tilde h}_2(t)={\tilde h}(t)$.
If we now define $\lambda(t)$ implicitly by
\beq
C(t,\lambda(t))=a^*
\label{lambda}
\eeq
$a^*$ the largest correlation in the scale, for large $t$ we have
\beq
\lim_{t \rightarrow \infty}{\tilde h}(t)-{\tilde h}(\lambda(t))=\ell(a^*)=0
\; .
\eeq
Defining $h(t)=\exp (-{\tilde h}(t))$
\beq
\jmath \circ C(t_{1},t_{2})=  \frac{h(t')}{h(t)}
\; .
\label{domi}
\eeq
with
\beq
\lim_{t \rightarrow \infty}   \frac{h(\lambda(t))}{h(t)}=a^*
\; .
\eeq

\newpage

{\large {\bf Figure Captions}}

\vspace{1cm}

\noindent {\bf Figure 1} A sketch of the function $f(a,a)$ {\it vs.} $a$,
$a \in [0,1]$.
\vspace{.5cm}

\noindent {\bf Figure 2} $\chi(t+t_w, t_w)$ {\it vs.} $C(t+t_w,t_w)$
The lines have been obtained
simulating a hypercubic spin glass cell of dimension $D=17$ at
a subcritical temperature $T=0.2$ for $t_w=30$, $100$, $300$, $1000$, $3000$.
The points represent
the static curve for the SK model.
\vspace{.5cm}

\noindent {\bf Figure 3}
Plot $C(t_{max},t)$ {\it vs} $C(t,t_w)$,
$t_w \leq t \leq t_{max}$, for
fixed $C(t_{max},t_w)=0.7$.  $D=15$ and $T=0.2$,
the four curves correspond to $t_w = 100$, $300$, $1000$ and
$3000$.
In the inset log-log plot, area {\it vs.} $t_w$,
$D=15$, $T=0.2$ and $t_w=30$, $100$, $300$, $1000$, $3000$.

\end{document}